\documentclass[conference]{IEEEtran}
\usepackage{float}
\usepackage[caption = false]{subfig}
\usepackage{amsmath,graphicx}
\usepackage{xcolor}
\usepackage{hyperref}
\usepackage{lipsum,multicol}
\usepackage{amsfonts}
\usepackage{ulem}
\usepackage{algorithm}
\usepackage{algorithmic}
\usepackage{babel}
\usepackage{amsthm}

\newtheorem{proposition1}{Proposition}

\graphicspath{ {img/} }
\title{A constrained Shannon-Fano entropy coder for image storage in synthetic DNA}
\author{
\IEEEauthorblockN{Xavier Pic, Marc Antonini}
\IEEEauthorblockA{
\textit{I3S laboratory, Côte d’Azur University and CNRS}\\
UMR 7271, Sophia Antipolis, France \\
xpic@i3s.unice.fr, am@i3s.unice.fr}
}
\begin{document}
%\ninept
%
\maketitle
\begin{abstract}
\par
The exponentially increasing demand for data storage has been facing more and more challenges during the past years. The energy costs that it represents are also increasing, and the availability of the storage hardware is not able to follow the storage demand's trend. The short lifespan of conventional storage media -10 to 20 years- forces the duplication of the hardware and worsens the situation. The majority of this storage demand concerns "cold" data, data very rarely accessed but that has to be kept for long periods of time. 
% \sout{Synthetic molecules, and particularly synthetic DNA are now becoming good candidates as alternatives for data storage \cite{DNARAM}. DNA molecules are composed of a sequence of nucleotides - adenine (A), cytosine (C), guanine (G) and thymine (T) - that can be used in synthetic DNA to represent data.} 
The coding abilities of synthetic DNA, and its long durability (several hundred years), make it a serious candidate as an alternative storage media for "cold" data.
% \sout{The Jpeg DNA\footnote{\color{blue}\url{https://jpeg.org/jpegdna/index.html}} research group has been developing softwares using the technologies of DNA synthesis to compress and store images into DNA. Some compression algorithm and coders have been developed specifically for this paradigm of data storage. }
In this paper, we propose a variable-length coding algorithm adapted to DNA data storage with improved performance. The proposed algorithm is based on a modified Shannon-Fano code that respects some biochemichal constraints imposed by the synthesis chemistry.
% \sout{that shows improved performance in terms of compression over the previous one, the Goldman coder \cite{Goldman2013}.} 
We have inserted this code in a JPEG compression algorithm adapted to DNA image storage \cite{Dimopoulou} and we highlighted an improvement of the compression ratio ranging from 0.5 up to 2 bits per nucleotide compared to the state-of-the-art solution, without affecting the reconstruction quality.
\end{abstract}
%
% \vspace{-1\baselineskip}
\section{Introduction}
Our society is relying more and more on the digital world, and as a consequence, the demand for digital data storage are exponentially increasing. The duration for which this data should be preserved is also constantly growing, mainly for legal reasons. The current tools are not sufficient for both this reasons, first because the storage hardware cannot follow the storage demand anymore, and secondly, because the hardware replication needed to ensure the durability of the data storage is confronted with the same demand problem. For those reasons, alternatives to the current data storage technologies have to be developed quickly. Synthetic molecules, and particularly synthetic DNA are promising candidates as alternative data storage media \cite{DNARAM}. Synthetic DNA molecules have huge encoding abilities (non-synthetic DNA is encoding living beings), are very small, and can be stored for very long periods of time.
%Thus, synthetic DNA is a good candidate for an alternative data storage medium.}
The DNA molecules are composed of a sequence of nucleotides - adenine (A), cytosine (C), guanine (G) and thymine (T) - that can be used to synthesize DNA and store digital data into it \cite{Grass}, \cite{Erlich}. 
% \sout{in synthetic DNA to represent data}.
Information theorists started to develop tools to encode data into DNA and some of them focused on image data storage \cite{Dimopoulou, Milenkovic}.
%{\color{red} IL FAUDRAIT CITER ICI UN ARTICLE DE MILENKOVIC}.
Recently, the JPEG DNA\footnote{\color{blue}\url{https://jpeg.org/jpegdna/index.html}} research group started an exploration on the opportunity to develop a specific image coding solution adapted to the technologies of DNA synthesis to compress and store images into DNA-like sequences \cite{antonini:hal-03447138}. 
% {\color{blue} NOT REALLY, YOU DON'T USE THE SYNTHESIS TECHNOLOGY TO COMPRESS, YOU DEVELOP YOUR OWN ECODING ALGORITHM FOR THAT EVEN IF IT'S CONDITIONED BY THE SYNTEHSIS METHOD THAT YOU'LL USE}
Some compression algorithm and coders have been developed specifically for this paradigm of data storage \cite{Goldman2013, DNAcoding, Milenkovic}.
\par
In this work we propose a variable-length quaternary encoder adapted to DNA data storage. This encoder is based on the Shannon-Fano algorithm revisited to work with a quaternary alphabet composed by the four letters (nucleotides) $\{A,C,G,T\}$, and constrained by some biochemical restrictions. In this paper we considered mainly the constraint on homopolymers which consists in avoiding codewords constructed by the repetition of the same nucleotide more than 3 times, e.g., $AAAAA$. This algorithm showed good performance compared to one of the best state-of-the-art coder proposed in \cite{Goldman2013}.
%The {\color{violet} proposed} variable-length coder 
% {\color{violet} WHICH ONE, THE ONE THAT YOU PROPOSE? BE MORE SPECIFIC} 
%that was adapted to fit DNA data storage problematics showed improvements in terms of compression over the previous one, the Goldman coder \cite{Goldman2013}.
Furthermore, we integrated the proposed quaternary coder in the image coding standard JPEG. The modified JPEG algorithm has shown better performance when compared to the image codec proposed in \cite{Dimopoulou} which uses the Goldman coder of \cite{Goldman2013}. We finally established bounds to identify the margin of progress still available in the proposed variable-length coder, to identify the improvement that could benefit the proposed DNA-based JPEG codec.
% {\color{violet} IF YOU ARE GONNA TO PRESENT YOUR WORK IN THE INTRODUCTION IT HAS TO BE CLEAR THAT IT'S YOUR WORK!!! ALSO, IF YOU ARE GOING TO COMPARE WITH GOLDMAN MAYBE YOU NEED TO EXPLAIN ROUGHLY WHAT IS IT}
% \vspace{-0.5\baselineskip}
\section{Context}
\subsection{DNA data storage}
%Our society is relying more and more on the digital world, and as a consequence, the demand for digital data storage are exponentially increasing. The duration for which this data should be preserved is also constantly growing, mainly for legal reasons. The current tools are not sufficient for both this reasons, first because the storage hardware cannot follow the storage demand anymore, and secondly, because the hardware replication needed to ensure the durability of the data storage is confronted with the same demand problem. For those reasons, alternatives to the current data storage technologies have to be developed quickly. Synthetic DNA molecules have huge encoding abilities (non-synthetic DNA is encoding living beings), are very small, and can be stored for very long periods of time. Thus, synthetic DNA is a good candidate for an alternative data storage medium. 
\par
A workflow has been designed to store data into those DNA molecules. The first step in this workflow is the construction of a coder able to represent data in a sequence of symbols of the alphabet $\{A,C,G,T\}$. The encoded information has to be input into a DNA molecule, through the operation of synthesis. The molecules then are stored in hermetic capsules. To access the data, the last operations of the workflow are used: first the DNA molecule is sequenced, which means that we read the sequence of nucleotides from the molecule. Then this nucleotide sequence has to be decoded to retrieve the data. In some of those processes, (synthesis, sequencing and storage), the DNA molecules are either manipulated or stored. During those events, a lot of factors can alter the integrity of the molecule and introduce errors in the nucleotide sequence. Those errors can be {\sc substitutions} (a nucleotide is changed into another one, for example an $A$ becomes a $G$) and {\sc indels} (random insertions and deletions of nucleotides in the sequence).
\par
Furthermore, to reduce the noise level of these different operations, the DNA coded information stream must respect some biochemical constraints on the combinations of bases that form a DNA fragment: homopolymers, high/low $GC$ content and repeated patterns should be avoided.
Designing a coding algorithm specific to synthetic DNA data storage appeared as a solution to accurately tackle the issues that come with the DNA storage workflow, particularly the noise produced in the workflow. Designing compression algorithms specific to DNA might also optimize the compression rate at which we store the data, which is still an important topic because of the high cost of the DNA synthesis.
\subsection{Image coding into DNA: A solution based on JPEG}
\begin{figure}
    \centering
    \includegraphics[width=9cm]{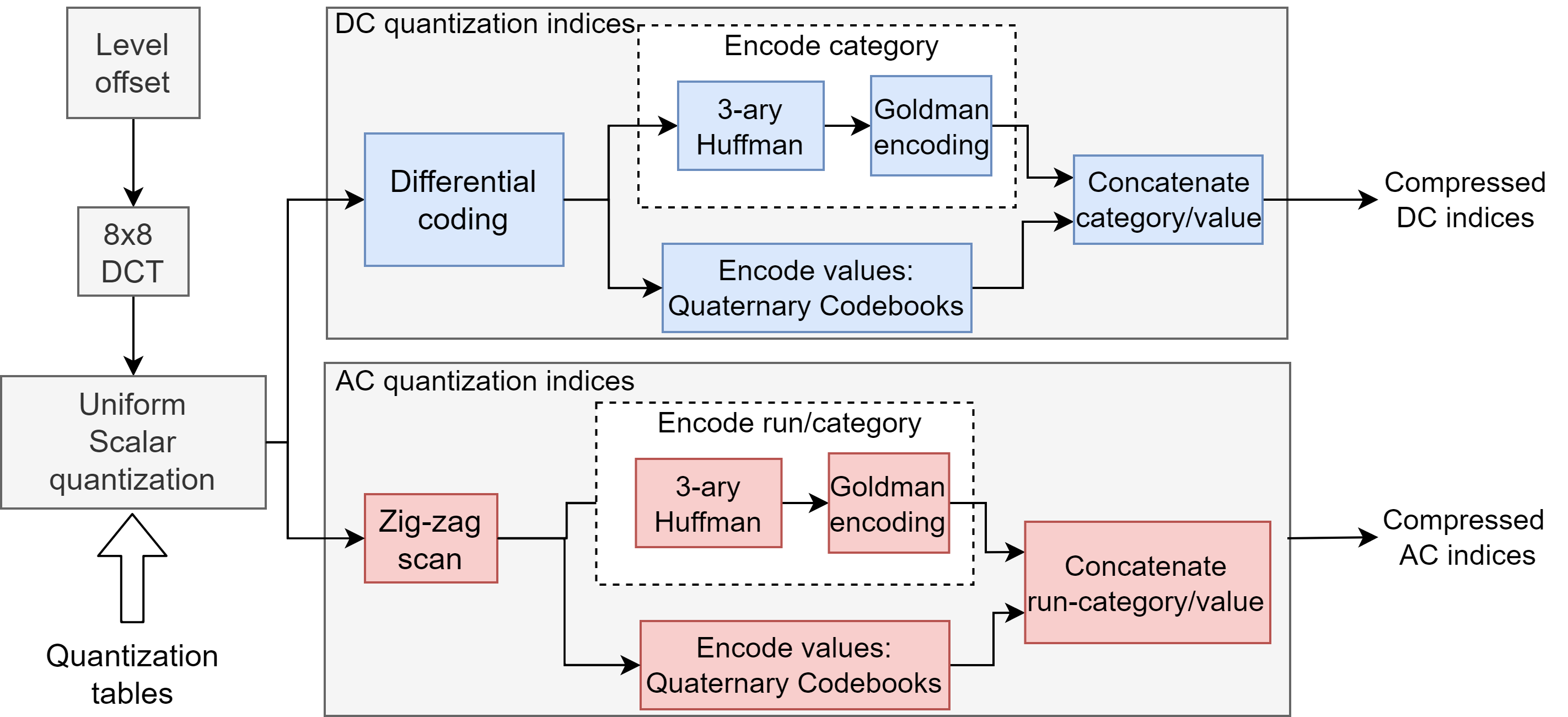}
    \vspace{-1\baselineskip}
    \caption{Compression workflow for image coding in a quaternary representation based on JPEG \cite{Dimopoulou,antonini:hal-03447138}.}
    \label{fig:jpegdna_workflow}
    \vspace{-1.5\baselineskip}
\end{figure}

The codec proposed in \cite{Dimopoulou} is a modified version of the classic JPEG standard algorithm, adapted to encode data into DNA. The classic JPEG algorithm compresses and encodes images with the use of two main encoders. Those encoders are used to encode the DC and AC indices obtained through the Discrete Cosine Transform (DCT) computed on the image, block by block. The first one is the Huffman coding that encodes the run/categories of the AC indices, and the categories of the DC indices obtained through the DCT. In \cite{Dimopoulou}, this coder has been replaced by the Goldman encoder \cite{Goldman2013} that generates, instead of binary codes, quaternary DNA-based codes. This work focuses on improving this specific coder to increase the performance of the general algorithm. The second coder used in the JPEG standard is a fixed-length coder that codes the values of the DC and AC indices obtained through the DCT. In \cite{Dimopoulou}, this coder has been replaced by the fixed-length quaternary coder proposed in \cite{DNAcoding}.
%the focus of different studies in the past and shows good performance in terms of coding capacity.
The figure \ref{fig:jpegdna_workflow} presents the whole workflow of the JPEG image coder adapted to DNA coding that has been proposed in our previous work \cite{Dimopoulou}.

In the rest of the paper, we focus our work on the design of a new efficient constrained quaternary variable-length encoder able to compete with the algorithm proposed in \cite{Goldman2013} and used in \cite{Dimopoulou}.
% \vspace{-0.5\baselineskip}
\section{Huffman Tree Coding}
\subsection{Background}
Entropy coding methods are variable-length lossless compression methods that represent the different symbols of the source it encodes. The codes are estimated with the help of a frequency table. For every symbol that needs to be encoded, the number of appearances of this symbol in the source is computed. Finally, the length of the codes representing every symbol varies with its probability of appearance: if a symbol is more frequent, the associated code will be short, if it is less frequent, the code will be long.
\begin{table}[h]
    \vspace{-0.5\baselineskip}
    \caption{Example of a binary Huffman code}
    \centering
    % \resizebox{4cm}{!}{
    \begin{tabular}{|c|c|c|}
        \hline
        Symbol & Probability & Code \\
        \hline
        a & 0.40 & 0 \\
        b & 0.05 & 1010 \\
        c & 0.18 & 110 \\
        d & 0.07 & 1011 \\
        e & 0.20 & 111 \\
        f & 0.10 & 100 \\
        \hline
    \end{tabular}
    % }
    \label{tab:huffman_example}
    \vspace{-0.5\baselineskip}
\end{table}
%Parler du code prefixe
\par
The Huffman algorithm is the optimal  symbol-by-symbol entropy coding method. Some theoretical limits on its performance have been established with the help of Kraft's inequality. 
If the source $S$ is represented by a random variable $X$ with a probability distribution $p$, the expected length $L$ of the Huffman code $C$ generated by this source can be defined as:
\begin{equation}
    L(C) = \sum_{x\in\Omega} p(X=x)l(C_x)
    \label{eq:expected_length}
\end{equation}
$C_x$ being the codeword representing the symbol $x$, $l(C_x)$ the length of this codeword and $\Omega$ the set of all possible symbols.
Given $b$ the base in which the Huffman algorithm encodes the source, the source's Shannon entropy $H_b(X)$ can be defined as:
\begin{equation}
    H_b(X) = \sum_{x\in X} -p(X=x) log_b(p(X=x))
    \label{eq:entropy}
\end{equation}
Shannon's source coding theorem \cite{Shannon} showed that, like for any other source code, the expected length of the $b$-ary Huffman code $C_b$ representing the source $S$ of random variable $X$ follows the law:
\begin{equation}
    H_b(X) \leq L(C_b) < H_b(X) + 1
\end{equation}
\subsection{The Goldman coder}
Nick Goldman introduced in \cite{Goldman2013} an entropy coding method adapted to DNA. This algorithm satisfies one of the biochemical constraints presented earlier: it doesn't generate codes with homopolymers. Homopolymers are the repetition of the same nucleotide base several times in a code or in a code stream (for example $AAAAA$ or $TTTTT$). 
To avoid those error-prone patterns, the Goldman code is generated by combining a ternary Huffman code and a rotating quaternary transcoder. The ternary Huffman coding generates codes using the alphabet \{0,1,2\}. The rotating transcoder translates this $\{0,1,2\}$ sequence into a quaternary sequence on the alphabet $\{A,C,G,T\}$, without reusing the same element twice in a row. Thanks to this rotating transcoder, no homopolymers will be generated by the Goldman coder.
\begin{table}[t!]
    \caption{Goldman base table for converting ternary Huffman symbols into nucleotides}
    \centering
    \resizebox{4.5cm}{!}{
    \begin{tabular}{|c|c|c|c|}
        \hline
        Previous nucleotide & 0 & 1 & 2 \\
        \hline
        A & T & C & G \\
        T & A & C & G \\
        C & A & T & G \\
        G & A & T & C \\
        \hline
    \end{tabular}
    }
    \label{tab:Goldman_table}
    \vspace{-1\baselineskip}
    % \vspace{-1\baselineskip}
\end{table}
The rotating transcoder doesn't affect the length of the codes generated by Huffman since it transcodes the source ternary symbol per ternary symbol. For that reason, the length of the codes for every symbol, and the expected length of the final code, are only conditioned by the Huffman coder. Hence, we can establish a boundary to the expected length of the entire Goldman coder $C_G$:
\begin{equation}
    H_4(X) \leq H_3(X) \leq L(C_G)
\end{equation}
\subsection{Limitations for constrained quaternary codes}
As previously mentioned, the theory has shown that the Huffman tree coding algorithm was the best entropy coding algorithm when source symbols are independent and identically distributed, {\it i.i.d.} source. This means that the quaternary Huffman tree coding algorithm will give the best performance in quaternary entropy coding of a {\it i.i.d.} source. The algorithm in itself is the theoretical limit for any quaternary entropy coding, hence the quaternary Huffman code $C_{H4}$ is a bound for any quaternary entropy coder $C$:
\begin{equation}
    \forall C \in Ce_4, H_4(X) \leq L(C_{H4}) \leq L(C)
\end{equation}    
$Ce_4$ being the set of all the possible quaternary entropy coders. The problem for DNA data storage is that, as we said previously, an unconstrained code will generate a lot of noise. This was the reasoning behind the development of the Goldman algorithm, and is still a problem that has to be considered. Thus, one cannot simply use the quaternary Huffman coder. This means that to develop a new entropy coder for DNA with improved performance over the Goldman algorithm or any ternary entropy coder $C$, a good objective in terms of performance would be to design a code $C$ such as:
\begin{equation}
    L(C_{H4}) \leq L(C) \leq H_3(X)
    \label{eq:objective}
\end{equation}
% \vspace{-0.5\baselineskip}
\section{Proposed variable length quaternary encoder}
\subsection{Motivations}
\label{subsection:motivations}
The goal of the Goldman coder \cite{Goldman2013} is to constrain the code and avoid the apparition of homopolymers. The experiments have shown that the biochemical processes of DNA data storage (synthesis, storage and sequencing) can tolerate homopolymers until a certain length (3 for Illumina sequencers and around 5 for Nanopore sequencers), otherwise, the noise starts increasing dramatically \cite{Ross}.
This means that the Goldman coder, that does not allow two consecutive identical nucleotides, is too restrictive when avoiding homopolymers.

%In order to meet the homopolymer constraint without losing too much coding potential, one could try to constrain the code to ternary only after having met the same nucleotide several times in a {\color{red} code}. However, the ternary tree-based Goldman algorithm is not a good candidate for {\color{red} designing such a coder since bla bla bla DIRE POURQUOI}\\
%{\color{blue} and a complete quaternary tree coder would not give any constraint to avoid too long homopolymers} {\color{red} CELA ME SEMBLE EVIDENT => PHRASE A ENLEVER ?}.\\

In order to meet the homopolymer constraint without losing too much coding potential, we proposed a solution that consists in a quaternary tree-coder constrained at certain positions to avoid too long homopolymers. Since tree structures are used for the code construction, the most simple solution was to constraint the arity (the number of sons of each node) of the tree at certain depths. 
The classic Huffman algorithm builds the coding tree from the leaves up to the root, and does not hold any information concerning the final depth of the tree nor the final depth of any of the nodes in the tree. Designing our algorithm from the Huffman algorithm, where there is a local control of the tree structure, did not appear as the most efficient solution, especially for obtaining a balanced tree.\\
Instead, we based our solution on the classic {\sc Shannon-Fano} recursive algorithm which builds the coding tree from the root, meaning that the current depth can be propagated through the recursions.

% \begin{figure}[b]
% \caption{Structure of the encoding tree with $max_{hl}=3$}
% \end{figure}
\subsection{Proposed constrained tree coder}
\begin{figure}[h!]
    \centering
    \includegraphics[width=8.5cm]{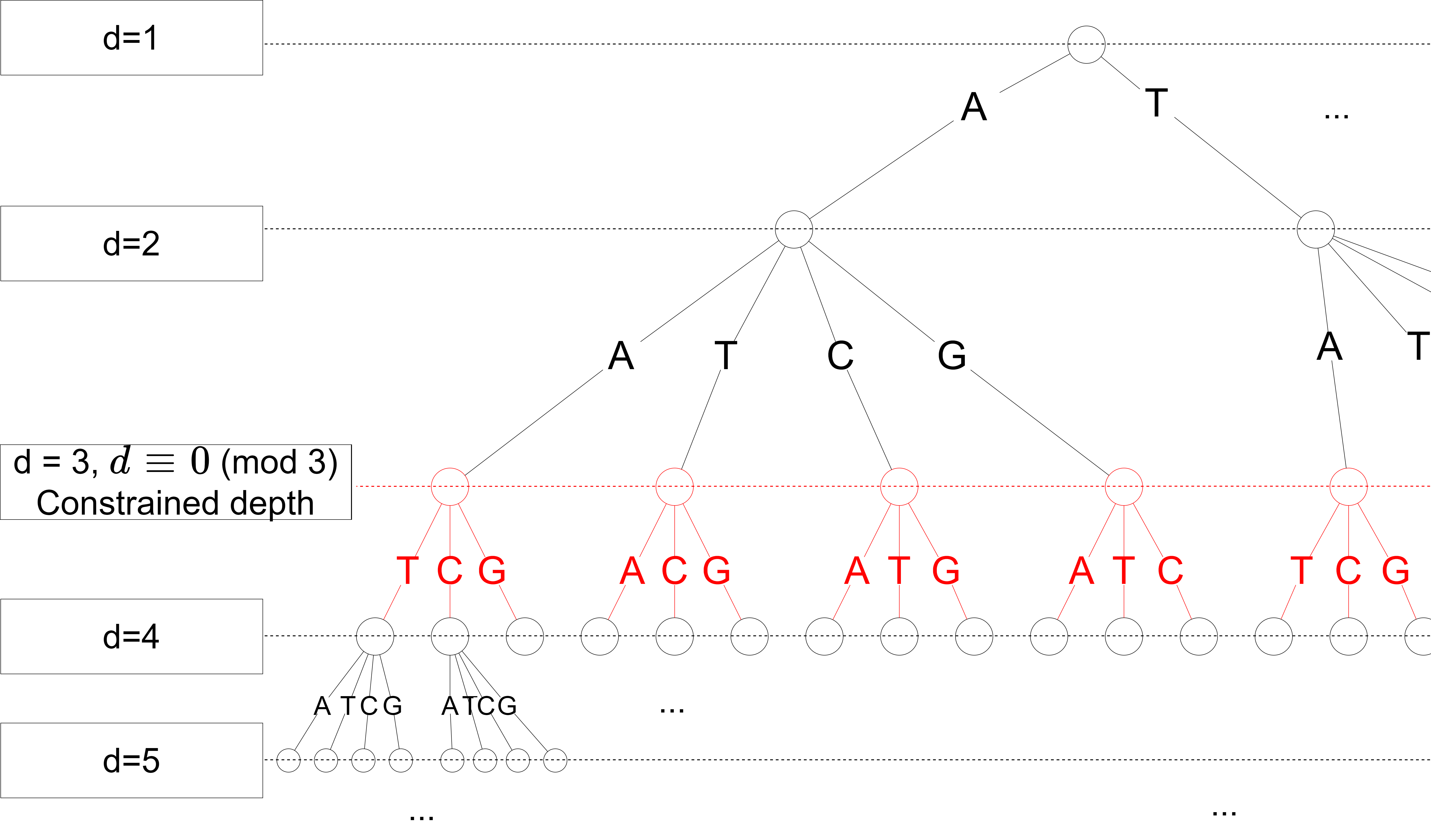}
    \vspace{-1\baselineskip}
    \caption{A section of a coding tree generated by the proposed algorithm for $max_hl = 3$, for readability, two of the top level branches and most branches at depth 4 to 5 have been deleted.}
    \vspace{-0.5\baselineskip}
    \label{fig:coding-tree}
\end{figure}

The coder is implemented with two algorithms. The algorithm \ref{algo:tree} is a simple initialization: it starts the whole construction of the tree, at the root (at depth 1).
The algorithm \ref{algo:tree_aux} manages the recursive calls in function of two parameters: a global one, the maximum length of homopolymer that is allowed $max_{hl}$, and a local one, the depth $d$ at which the call was made.
If the depth $d$ is in the class 0 of congruence modulo $max_{hl}$, the tree will be constrained to ternary, otherwise, it will not be constrained. The figure \ref{fig:coding-tree} shows an example of such a coding tree.

%{\color{red} EXPLIQUER CLAIREMENT LES ALGORITHMES! CE N'EST FAIT NUL PART. MONTRER LEUR FONCTIONNEMENT SUR UN EXEMPLE SIMPLE.}
In the algorithms \ref{algo:tree} and \ref{algo:tree_aux}, we assume that the symbols in the set $S$ and their frequency table $f$ have both already been sorted by %{\color{red}\sout{order of}} 
decreasing frequency. In the algorithm \ref{algo:tree_aux}, the function $partition(S, f, n)$ slices into $n$ divisions the set of symbols $S$ and frequencies $f$, with the cumulated frequencies of every slice being as close as possible to each other.
The function $merge\_trees(P)$ takes the list $P$ containing 3 or 4 trees and generates a tree with a new root, and each son of this root being one of the trees in the input sequence $P$. The first son is labeled with a 0, the second with a 1, the third with 2 and the last one with a 3.
\begin{algorithm}[h!]
\begin{algorithmic}
\caption{$Sfc(S, f, max_{hl})$: Create the DNA coding tree $T$ for a set of symbols $S$, $f$ their frequency table and $max_{hl}$ the maximum length of homopolymer that is allowed.}
\label{algo:tree}
\RETURN $Sfc\_aux(S, f, max_{hl},1)$
\end{algorithmic}
\end{algorithm}
\begin{algorithm}[h!]
\begin{algorithmic}
\caption{$Sfc\_aux(S, f, max_{hl}, d)$: Create the DNA coding tree at depth $d$.}
\label{algo:tree_aux}
\IF{$length(S) = 0$}
\RETURN $\emptyset$
\ELSIF{$length(S) = 1$}
\RETURN $create\_leaf(S)$
\ELSIF{$d \equiv 0 \pmod {max_{hl}}$}
\STATE $PS, Pf \leftarrow partition(S, f, 3)$   //Constrained case
\ELSE
\STATE $PS, Pf \leftarrow partition(S, f, 4)$   //Unconstrained case
\ENDIF
\STATE $PT \leftarrow$ empty list of length $length(PS)$
\FOR{$i \leftarrow 1$ to $length(PS)$}
\STATE $ PT[i] \leftarrow Sfc\_aux(PS[i], Pf[i], max_{hl}, d+1)$ 
\ENDFOR
\STATE $T \leftarrow merge\_trees(PT)$
\RETURN $T$
\end{algorithmic}
\end{algorithm}
\subsection{Transcoding into DNA}
The algorithm described earlier generates a code containing for every symbol a sequence of bases: 0, 1, 2, and 3. These sequences still need to be translated into DNA sequences respecting the homopolymer constraint described earlier.

To transcode the sequences into DNA, two different kinds of transcoding tables are used. According to the tree structure, the $k^{th}$ symbol in the sequence is a ternary symbol if $k \equiv 0 \pmod {max_{hl}}$ and a quaternary one otherwise, $max_{hl}$ being the maximum homopolymer length. When the symbol is ternary, it is translated into a nucleotide with the Goldman transcoding table  (Table \ref{tab:Goldman_table}), otherwise, the quaternary trancoding table (Table \ref{tab:quaternary_table}) is used.
%{\color{red} DANS QUELLE MESURE POURRAIT-ON AUTORISER LA TAILLE D'UN HOMOPOLYMERE ETRE SUPERIEURE A 3 AVEC CET ALGORITHME}. {\color{teal} En changeant la variable $max_hl$ qui vaut par défaut 3, on peut choisir une taille d}}{\color{red} MAIS ALORS COMMENT SON GÉRÉS LES TERNARY ET QUATERNARY HUFMAN ?Toujours de la meme façon: on met en ternary la dernière profondeur modulo la taille maximale autorisée des homopolymères (max_hl) détermine les profondeurs où on contraint en ternaire}

\begin{table}[h!]
    \vspace{-0.5\baselineskip}
    \caption{Quaternary transcoding table}
    \centering
    % \resizebox{4cm}{!}{
    \begin{tabular}{|c|c|c|c|c|}
        \hline
        Symbol & 0 & 1 & 2 & 3 \\
        Nucleotide & A & T & C & G \\
        \hline
    \end{tabular}
    % }
    \label{tab:quaternary_table}
    \vspace{-0.5\baselineskip}
\end{table}

\subsection{Property of the proposed tree coder}
If $max_{hl}$ is the maximum homopolymer length that can be accepted, constraining the coding tree to ternary at every depth that is a multiple of $max_{hl}$,
% \sout{would limit the apparition of homopolymers between two following ternary constrained depths, }
% {\color{red}\sout{meaning between two depths $d1$ and $d2$ that are consecutive multiples of $max_{hl}$, so $d2 = d1+max_{hl}$
% REPHRASE, NOT CLEAR}
% }
would prevent the apparition of homopolymers of length more than $max_{hl}$ (see Property \ref{proposition1}). 
\begin{proposition1}
\label{proposition1}
Let \(C_{SFC}\) be a code obtained thanks to the proposed coder (algorithm \ref{algo:tree}). Let $\mathcal{S}$\((C_x)\) be the set of homopolymer patterns found in the codeword \(C_x\) of the symbol \(x\). Let \(\omega_h \in \mathcal{S}(C_x)\) be a homopolymer subsection of \((C_x)\). Let \(d_i\), and respectively \(d_j\) be the depth at wich the section \(\omega_h\) starts, respectively ends, in the coding tree. \\Then, $\exists!{k} \in{\mathbf{N}}$ with,\\
\begin{equation}
    \begin{split}
        k \times max_{hl} &< d_i < (k+1) \times max_{hl}\\
        k \times max_{hl} &< d_j \leq (k+1) \times max_{hl}
    \end{split}
\end{equation}
\end{proposition1}
\begin{proof}
By contradiction: If a code $C_{SFC}$ has a codeword $C_x$ containing a homopolymer pattern $\omega_h$ of length bigger than $max_{hl}$, the homopolymer can be located in the coding tree between depths $d_i$ and $d_j$ with $l(\omega_h) = d_j-d_i > max_{hl}$. By construction of the coding tree, there is a constrained depth in the middle of the homopolymer pattern, where the Goldman transcoding table \ref{tab:Goldman_table}
%{\color{red} Goldman transcoder QUE VIENT FAIRE GOLDMAN ICI ?}{\color{teal} Le système d'étiquetage de Goldman est utilisé au niveau des profondeurs contraintes pour éviter les homopoolymères, peut-être vaut-il mieux dire Goldman transcoding table en faisant référence à la table qu'il y a déjà dans le papier} {\color{red} PROBABLEMENT...} 
is used, hence two consecutive nucleotides are not identical, so $\omega_h$ is not a homopolymer pattern.
\end{proof}
% \vspace{-0.5\baselineskip}

%{\color{red} JE N'AI PAS LU LA SUITE MAIS :\\
%(I) JE NE COMPRENDS PAS POURQUOI IL Y A DEUX PARTIE EXPERIMENTALES? CA NE SE FAIT PAS...\\ (II) IL NE FAUT PAS APPELER LE CODEC JPEG DNA SINON CERTAINS VONT GRINCER DES DENTS !!! ET EN PLUS LE LECTEUR SERA PERDU DANS LA COMPREHENSION...}
\section{Experimental results}
%The final goal of this paper is to improve the performance in terms of compression of the JPEG DNA algorithm. A final experiment would be to compare the performance of the usual JPEG DNA algorithm, observe the compression rates obtained, then, integrate the new variable-length coding algorithm and analyze the gain - or loss - that it gives for the same parameters of compression. 
% {\color{violet} This type of experiment will show us if for this problem of image compression, the new algorithm works better or not, but it will not allow us to quantify, or at least to have an estimate, of the improvement the variable-length coder has in itself. }
% {\color{violet}TRY TO WRITE THIS IN A MORE PROFESSIONAL/ SCIENTIFIC WAY, I CAN TRY TO HELP YOU WITH THAT TOMORROW}
%{\color{blue} Although this type of experiment can show the improvement of the new algorithm over the previous one, quantifying, or at least estimating the gains produced on the variable length coder itself require specific preliminary experiments conducted on the coder only, and not on the full image codec.}
\subsection{Experiments on synthetic sources}
Before integrating the coder inside a complete image codec, we estimate the raw gain that we can obtain in comparison to the Goldman coder ($C_G$).
We also compare our results to a Huffman-based constrained quaternary code ($C_{H4C}$) that we implemented, even if as mentionned in the motivations subsection (\ref{subsection:motivations}) it was not considered as a good candidate from the beginning.
\par
The metric used to compare the performance of variable-length coders is the expected length, that has been previously introduced in equation (\ref{eq:expected_length}). 
%It is simply a probability-weighed average over the length of the codes representing each symbol. 
\par
We used {\it i.i.d} Gaussian sources for those experiments. It was made sure to generate enough samples in the sources to approach the Gaussian distribution: 100 relisations, each one containing 10000 samples.
%{\color{red}
Each sample is quantized using a dictionnary composed of 162 symbols.
%(?) COMBIEN DE REALISATION DE LA SOURCE ON ÉTÉ FAITES PAR SIMULATION ?}. 
Since the Goldman coder is based on a Huffman ternary tree coder, comparing the expected length of the two coders to the ternary entropy (equation (\ref{eq:entropy}) with $b=3$) is appropriate: if the new coder beats the ternary entropy, then it is better than any tenary-based entropy coder for DNA (see equation (\ref{eq:objective})).\\
\begin{table}[h!]
    \vspace{-0.5\baselineskip}
    \caption{Length expectancy results of the entropy coders, expressed in nucleotides per symbol, on Gaussian sources, with $max_{hl} = 3$.}
    \centering
    % \resizebox{8.75cm}{!}{
    \begin{tabular}{|c|c|c|c|c|c|}
    \hline
        $H_4(X)$ & $L(C_{H4})$ & $\boldsymbol{L(C_{SFC})}$ & $L(C_{H4C})$ & $H_3(X)$ & $L(C_G)$\\
        3.48 & 3.56 & \textbf{3.81} & 4.31 & 4.39 & 4.45\\  
        \hline
    \end{tabular}
    % }
    \label{tab:gaussian_results}
    \vspace{-0.5\baselineskip}
\end{table}

The length expectancy presented in table \ref{tab:gaussian_results} shows a gain of 0.64 nucleotide per symbol for the proposed algorithm ($C_{SFC}$) over Goldman ($C_G$), and a difference of 0.25 nucleotide per symbol with the unconstrained quaternary Huffman coder ($C_{H4}$). As expected in the motivations subsection (\ref{subsection:motivations}), in comparison, the constrained Huffman-based coder $C_{H4C}$ clearly underperforms. Not having a control over the depth of the coding tree means that it is harder to balance it and optimize the expected length.
\par
As a last experiment, the performance of the coder was evaluated on a source following a distribution that is similar to the distribution of the AC indices produced by a JPEG codec in most natural images.
The results presented in the table \ref{tab:freq_results} have been generated using a source that follows a frequency table of the AC coefficients computed on the Kodak image dataset\footnote{\color{blue}\url{http://r0k.us/graphics/kodak/}}. Here again, the proposed constrained Shannon-Fano quaternary encoder is the best performing.
\begin{table}[h]
    \caption{Length expectancy results of the entropy coders, expressed in nucleotides per symbol, on sources using AC values frequencies, with $max_{hl} = 3$}
    \centering
    % \resizebox{8.75cm}{!}{
    \begin{tabular}{|c|c|c|c|c|c|}
    \hline
        $H_4(X)$ & $L(C_{H4})$ & $\boldsymbol{L(C_{SFC})}$ & $H_3(X)$ & $L(C_{H4C})$ & $L(C_G)$\\ 
        1.21 & 1.39 & \textbf{1.43} & 1.53 & 1.61 & 1.64\\
        \hline
    \end{tabular}
    % }
    \label{tab:freq_results}
    \vspace{-1\baselineskip}
\end{table}
\subsection{Experiments on image coding}
As explained in the introduction, the variable length coder is only one of the coders used in the JPEG DNA experimentation software's algorithm \cite{antonini:hal-03447138}. It is used to encode the category of a DC coefficient or the run/category (a combination of the AC category and of the run-length-encoded sequence of zeros preceding this AC coefficient) of an AC coefficient. In the classic JPEG DNA experimentation software's algorithm, the Goldman encoder encodes those values. To study if and how the proposed variable length coding algorithm improves the compression, we need to make it so that during the comparison, the only variant is the choice of the variable length coder for the categories and run/categories.
\par
\begin{figure}[h]
    \centering
    \includegraphics[width=9cm]{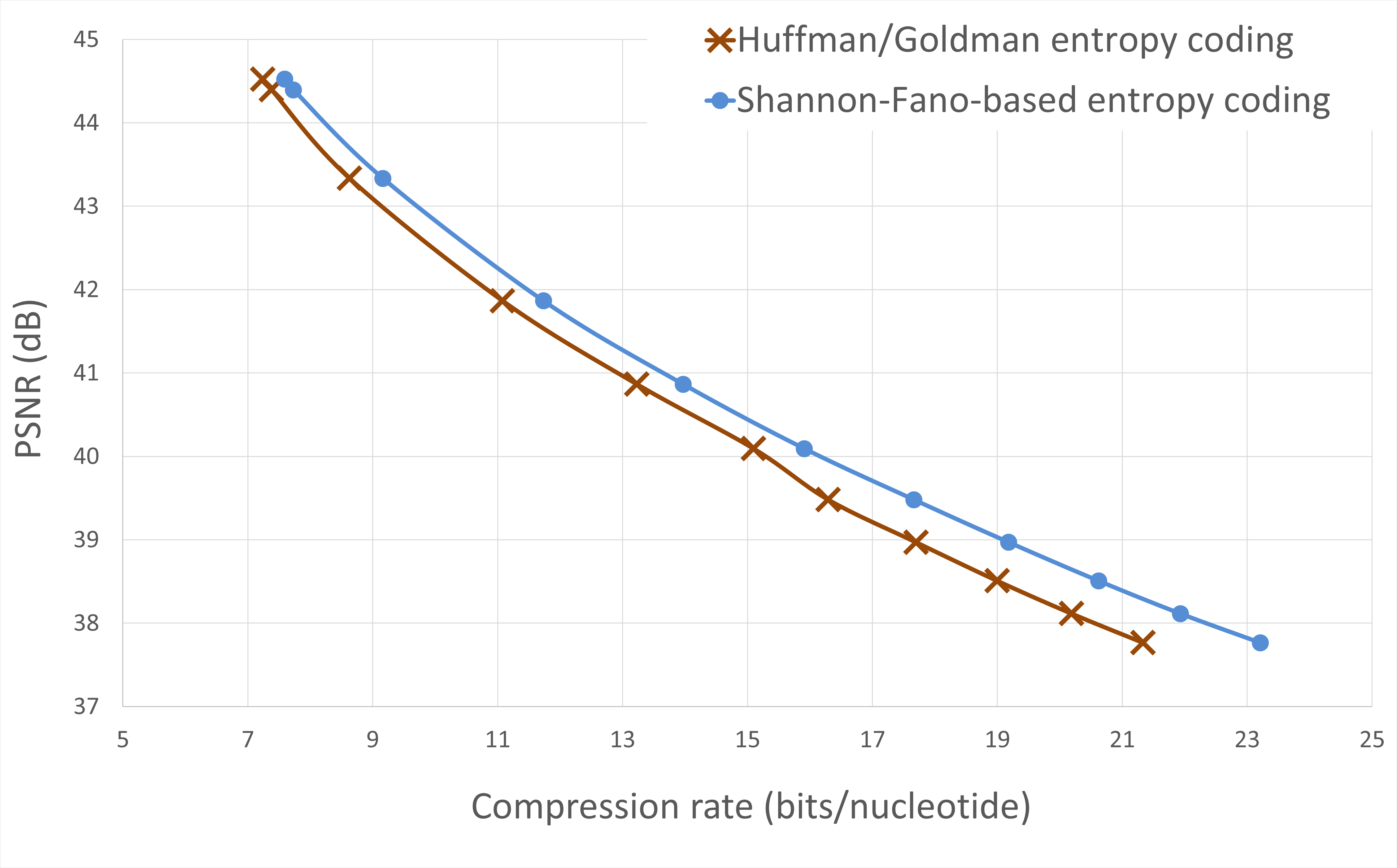}
    \vspace{-1.5\baselineskip}
    \caption{Compression performance comparison for the image kodim23 from the kodak dataset between the proposed quaternary encoder and the Goldman encoder. Both encoder have been implemented in the JPEG DNA experimental software, with $max_{hl} = 3$}
    \vspace{-1.5\baselineskip}
    \label{fig:comparison}
\end{figure}

To evaluate the performance of the two codecs we computed the compression ratio, expressed in bits per nucleotide. It corresponds in our case to the ratio of the size of the input image (in bits) over the total size of the DNA sequences (i.e., the number of nucleotides) generated to compress the image. This compression ratio is evaluated for the same quality parameters for the two different codecs. The PSNR in function of the compression ratio is presented in the figure \ref{fig:comparison}. The kodak dataset was used to evaluate the performance of the new image compression algorithm.
\par
Through all the images in the test image dataset, our new variable-length coder has consistently improved the performance of the compression algorithm over the one using the Goldman entropy coder. For example with the image kodim23 from the kodak dataset, the compression rate gain varies between 0.5 and more than 2 bits per nucleotide as shown in the figure \ref{fig:comparison}. 
The relative gain is less important than when using the entropy coder as a standalone coder because in the case of the image codec, other coders are used, and those have not been optimized.
\par
%The margin of improvement of the entropy coder for the image codec is quite narrow {\color{red} ???}{\color{teal}:} with a source that follows the distribution usually found in the frequency tables for JPEG, the proposed method's expected length is 0.04 nucleotide per symbol longer that the unconstrained quaternary Huffman code ($L(C_{H4})$), the Goldman coder is 0.25 nucleotide per symbol longer.
% \vspace{-0.5\baselineskip}
\section{Conclusion}
We have developed a new variable-length tree coder adapted to DNA data storage. This new coder has shown improvements over the Goldman entropy coder that is usually chosen in variable length DNA coding problems. We have even shown that this new coder would do better than any ternary entropy coder by having an expected lenth inferior to the ternary entropy of the source.
We then have integrated this new entropy coder inside an image codec to analyze its influence on the performance of the codec. We have seen consistent improvements on the compression rate with the dataset that we used for testing. 
\par
In the future, it would be interesting to study the resilience of this new encoder to the noise produced by the different processes of the DNA storage workflow, namely the synthesis, storage and sequencing. Developing new consensus algorithm, or maybe even error correction codes could also bring a lot to this new coder. Finally, it could be interesting to mathematically establish theoretical bounds to the performance of the proposed coder.
% \vspace{-0.5\baselineskip}

\end{document}